# Tuition too high? Blame competition

Oleg V. Pavlov [a],*, Evangelos Katsamakas [b]

[a] *Department of Social Science and Policy Studies, Worcester Polytechnic Institute, Worcester, MA 01609, opavlov@wpi.edu*
[b] *Gabelli School of Business, Fordham University, New York, NY 10023, katsamakas@fordham.edu*
* *Corresponding author*

**Abstract**
In this article, we develop a feedback theory that includes reinforcing and balancing feedback effects that emerge when colleges compete for reputation, applicants, and tuition revenue. The feedback theory is replicated in a formal duopoly model consisting of two competing colleges. An independent ranking entity determines the relative order of the colleges. College applicants choose between the two colleges based on the rankings and the financial aid offered by the colleges. Contrary to the conventional wisdom that competition lowers prices and benefits consumers, our simulations show that competition between academic institutions for resources and reputation leads to tuition escalation that negatively affects students and their families. Four of the five scenarios – rankings, a capital campaign, facilities improvements, and an excellence campaign – increase college tuition, institutional debt, and expenditures per student; only the scenario of ignoring the rankings decreases these measures. By referring to the feedback structure of academic competition, the article makes several recommendations for controlling tuition inflation. This article contributes to the literature on the economics of higher education and illustrates the value of feedback economics in developing economic theory.

## 1  Introduction

The highly competitive nature of the American system of higher education has established many American universities as the most prestigious and admired institutions in the world (MacLeod and Urquiola, 2021). While students compete for admissions to the best schools and faculty compete for academic positions, higher education institutions, in their pursuit of excellence, compete for students, faculty, research funding, and donations. For years, academic competition has been seen as an effective sorting and resource allocation mechanism for directing the best students, the most productive faculty, and plentiful resources to the best universities.

Being free to invest in incentives that attract faculty and students, academic institutions are engaged in positional competition that compels even less selective universities to spend a great deal of resources on education-related activities as well as non-educational services including dining, housing, hotels, and athletics (Frank, 2004; Vedder, 2019; McClure, 2019). Because excessive academic rivalry might be harmful to universities, there is a strong need for a better understanding of the negative effects of academic competition with a special focus on the sustainability of the entire educational sector (Frank, 2004; McClure, 2019; Naidoo, 2016; Reuben, 2022). Our analysis addresses this research need.  The article contributes to the emerging literature on the competition between higher education institutions that has not yet been thoroughly studied (Jacob et al., 2018).

Earlier research has suggested a conceptual framework that explains that positional rivalry between colleges and universities generates strong feedback effects that encourage academic excellence (MacLeod and Urquiola, 2021) while also increasing educational costs and tuition (Frank, 2004; Winston, 2000). Such circular effects can be formally modeled using feedback economics (Cavana et al., 2021) which studies economic problems with the tools of system dynamics. System dynamics is a methodology that originated at MIT Sloan School of Management in the late 1950s (Forrester, 1958; Forrester, 1995) and has become a popular approach (Jahangirian et al., 2010) for models that incorporate feedback effects, complex causality, delays, and stock-and-flow relationships (e.g. Maani and Cavana, 2007; Morecroft, 2007; Sterman, 1985). The late Professor Richard Day, whose research legacy is celebrated in this special issue, advocated for using system dynamics in economic analysis (e.g., Day, 1974; Day, 1983c). Following Day's pioneering vision, we hope that this article will encourage even more economists to adopt feedback analysis.

The next section on feedback analysis presents the system dynamics approach, then discusses the contribution of Richard Day and reviews several feedback models developed for higher education. Section 3 revisits some prominent literature on the market for higher education, and offers some background on tuition escalation, college rankings and the non-educational competition between colleges. In this paper, unless we explicitly state otherwise, 'tuition' means 'sticker price'. In Section 4, we construct the feedback theory of competition between colleges. Section 5 introduces a formal duopoly model of two nonprofit tuition-based colleges that provide educational services and amenities to students who exhibit preferences for college rankings. The model is used in scenario experiments in Section 6 that confirm that competition between institutions for resources and reputation leads to surging tuition costs. The discussion in Section 7 suggests that to cope with the cost of college attendance, we must address the structural issues of academic competition. The concluding section reviews the limitations of this study and suggests directions for further research.

## 2  Feedback analysis

This article uses feedback economic analysis that has its roots in system dynamics. Cavana et al. (2021) provide a broad sample of feedback models and relevant background literature on the use of system dynamics for economic analysis. In this section, we briefly review system dynamics as well as early papers by Richard Day in which he advocated for using system dynamics in economic analysis. We will conclude this section with a recap of several feedback models of higher education.

## 2.1 System Dynamics

System dynamics is a methodology for modeling complex socioeconomic systems (Maani and Cavana, 2007; Morecroft, 2007). Mathematically, system dynamics models are systems of nonlinear and non-stochastic partial differential equations that are solved numerically. Such models typically include quantitative and qualitative variables which form causal chains and feedback loops that include delays. The fundamental idea of system dynamics is that system structure determines system behavior. Since its invention at the Sloan School of Management at MIT in the 1950s (Forrester, 1958; Forrester, 1995), system dynamics been successfully used to study numerous real world problems (e.g., Cavana, 2021; Hovmand, 2013; Lane et al., 2015; Moxnes, 2000; Morecroft, 2007; Maani and Cavana, 2007; Pavlov et al., 2008 ; Saeed, 1994; Saeed, 2022; Stephens et al., 2005; Vennix, 1996). The field has evolved into an academic discipline with peer-reviewed publications, an academic society, and educational programs (Bosch and Cavana, 2018; Davidsen et al., 2014; Pavlov et al., 2014; Schaffernicht and Groesser, 2016). While this feedback modeling approach is popular in many fields (Galbraith, 2010), its tools are not yet widely used by economists (Radzicki, 2021).

## 2.2 The legacy of Richard Day

The late Richard Day, whose research legacy we celebrate in this special issue, was an economics professor at University of Southern California and the founding co-editor of this journal with Sidney G. Winter. Day worked on models of adaptive economic processes (e.g., Day, 1982b; Day, 1983b; Day, 1994; Day, 2001). He believed that system dynamics models had great potential for investigating complex economic behavior because they are logical and relatively straightforward for those who are familiar with the system dynamics notation (Day, 1982a). In his first paper on system dynamics (Day, 1974), Day explained the system dynamics vocabulary and diagramming conventions. As an example, he translated the growth model developed by MIT economist Robert Solow into the system dynamics model and identified the feedback loops in Solow's model.

In his 1975 paper (Day and Koenig, 1975), Day reviews the World models developed by Jay Forrester and his team at MIT to study the interaction of natural resources, production, ecology and population. The models predicted a sudden economic decline in the 21$^{st}$ century that could be solved by such measures as zero population growth, recycling, and zero net capital formation. He agrees that at that time, World models were the only models that incorporated population, capital, resources, food, and pollution. He points that the World models provide an opportunity to study dynamic interaction of many variables because they were more complex than contemporary neoclassical growth models. Day reviews the advantages of using system dynamics modeling, including the intuitive nature of graphical representation of causality and logic of such models and the flexibility of using graphical functions to represent different relationships. Day argues that the traditional economic approach does not pay enough attention to the analysis of unintended consequences and tends to overemphasize the estimation of parameters. He notes that even if some parameters are guesses, the model is still useful for policy analysis. Day points out that *The General Theory* by Keynes, which was a very influential book, was completely qualitative. Similarly, Samuelson's *Principles of Economics* explains the concept of multiplier, accelerators, monopoly, and other economic concepts with very crude estimates of parameters.

Day's subsequent research refined the use of systems dynamic for economic theory building. In Day (1981), he developed a general theory of epoch switching driven by economic adaptation that explain such evolutionary events as switching from hunting to agricultural production. He discussed the complex causality and feedback effects with graphs borrowed from the system dynamics toolset and constructed a computational model using the system dynamics modeling notation and software. Day (1982a) explored the suitability of the system dynamics method for representing nonlinear economic models that could generate complex economic behavior. He created a system dynamics model that shows a causal loop diagram and the flow diagram for Lorenz's chaos equations and demonstrates this model's ability to generate chaotic trajectories. He wanted to use system dynamics models to contribute to the endogenous theory of economic irregularity and posited that the emphasis on counterintuitive behavior observed in nonlinear economic models can be studied with system dynamics (Day, 1982a).

Day understood that the work of the system dynamics researchers was kindred in spirit to his work (Day, 1983a). Day recognized that "simulation models constructed according to the Forresterian paradigm" also study endogenous nonlinear relationships in economic systems that produce unexpected behavior. He noticed that the system dynamics research field started by Jay Forrester shared the same general goals of studying socioeconomic dynamics and observed many parallels between his work and work in the system dynamics field (Day, 1983c). In the spirit of Richard Day's research, we use a system dynamics approach to develop a feedback model to demonstrate that competition between colleges that vie for reputation and resources leads to tuition inflation.

## 2.3   Feedback models of higher education

Because problems in higher education are complex and dynamic, researchers have advocated for studying them using feedback systems approaches (e.g., Barnabe, 2004; Bell et al., 2000; Kennedy and Clare, 1999; Massy, 2016; Oyo and Williams, 2009; Rouse, 2016). For example, Saeed (1996) studied conditions that led to the emergence of an authoritarian administration at a college in the context of a developing country. Galbraith (1998) analyzed the effect of national policies in Australia on university management. He identified generic causal structures that arise in the presence of competition between universities and internal institutional units for students and scarce resources. His analysis suggests that escalation scenarios, erosion of goals and the "tragedy of the commons" outcomes are likely to occur. Bergland et al. (1999) used computer experiments with a system dynamics model to forewarn the administration of a college of an upcoming budgetary collapse due to the perilous student admission policies.

Barlas and Diker (2000) built an interactive system dynamics model for long-term management of undergraduate and graduate enrollments, number of faculty, teaching quality, research output, and outside consulting calibrated for Boğaziçi University in Turkey. The model did not include budget planning. Using the model with stakeholders revealed that different measures were important to different stakeholders.

Strauss and Borenstein (2015) developed an aggregate system dynamics model of the undergraduate education in Brazil. The model was used to understand the difficulties faced by the country in achieving stated national educational goals. It tests scenarios linked to government regulation and distribution of private and public institutions.

Al Hallak *et al.* (2019) used feedback analysis to investigate student enrollment at private higher education institutions in Syria. Their study is at the university level and includes student enrollments, university rank, selectivity, and university financials. In 2003, the Syrian government allowed private universities to help public universities to meet the demand for education. The authors noted that private universities in Syria are less competitive than public universities. While stopping short of developing a computational model, their paper recommended investing in faculty and facilities to increase reputation of private institutions, while controlling tuition increases.

## 3 Background

This section provides the essential background on the market for higher education and reviews the problem of tuition inflation. Additionally, it covers college rankings and the facilities arms race, which have been recognized as contributors to the escalating costs in higher education.

### 3.1 Market for higher education

Hoxby (1997) analyzes changes in the structure of American college market due to geographic integration that occurred since 1940. Geographic integration happened for a variety of reasons such as the G.I. Bill after World War II, the introduction of standardized admissions testing, the introduction of standardized financial needs analysis, and the deregulation of the airlines and phone companies that reduced the cost of travel and communication. Before the integration of the market, colleges acted as autarkies. Geographic integration meant that more students could attend colleges out of state, and therefore integration increased competition between colleges for students, faculty, and administrators. Hoxby further argues that this type of competition led to more stratification of colleges on quality across the nation, but colleges themselves became more homogeneous in terms of student ability. This quality competition also led to higher college prices as most selective colleges can exercise lots of market power in price setting. Geographic integration led to more competition between colleges that resulted in better average quality, higher tuition, and higher expenses per student.

Epple et al. (2006) explain college stratification by student ability and household income. They adapt their earlier model of competition between schools (Epple and Romano, 1998) to develop a theoretical model of provision of undergraduate higher education. The model (Epple et al., 2006) includes potential students who are on a continuum of household income and academic ability that choose between several colleges. The authors assume that the quality of educational experience is a function of peer ability and income as well as instructional expenditure per student. The model treats tuition and college endowments as exogenous to the analysis. It predicts stratification of colleges by student ability and household income. The model demonstrates that a policy that provides resources to low-income families increases the likelihood that low-income students with high ability can attend elite colleges because higher income students can cross-subsidize the cost of education for these students.

Jacob et al. (2018) study the importance of consumption amenities in competition between colleges. The authors estimate a discrete choice model for student preferences for college attributes such as academic quality, net price, and consumption amenities. They found preference heterogeneity among high school seniors who were asked about the importance for college selection of academics, costs, peer quality, and non-educational services. Lower-achieving students are willing to pay more for consumption amenities and higher-achieving students have stronger preferences for academic quality. Due to this heterogeneity in student preferences, colleges face different preferences depending on the market segment that enrolls at their institution. The authors find that colleges respond more to the demand for amenities when students value them more that leads to heterogeneity in amenity provision among colleges.

## 3.2 Tuition escalation

Surging tuition cost has been a concern for over a century. For example, The New York Times on November 26, 1874 published an op-ed that included the following statement, "It is not encouraging to think of the coming of a time when college training will be so costly as to confine it to the sons of well-to-do men" (Op-ed, 1874). Tuition continued to grow rapidly in the 20$^{th}$ century. In the late 1960s, Bowen (1968) showed that tuition at selective private universities had increased by 2-3 percentage points above inflation during 1905-1965. In the mid-1990s, Massy (1996: 20) observed that tuition grew at rates of about 3 to 5 percent per year when adjusted for inflation in the period between 1975-1990. Figure 1 extends those observations to the present day by plotting the average published tuition, fees, room, and board data for private nonprofit four-year institutions against the Consumer Price Index (CPI) between 1972 and 2020, keeping 1972 as the base year. As can be seen in the figure, tuition still grows faster than inflation.

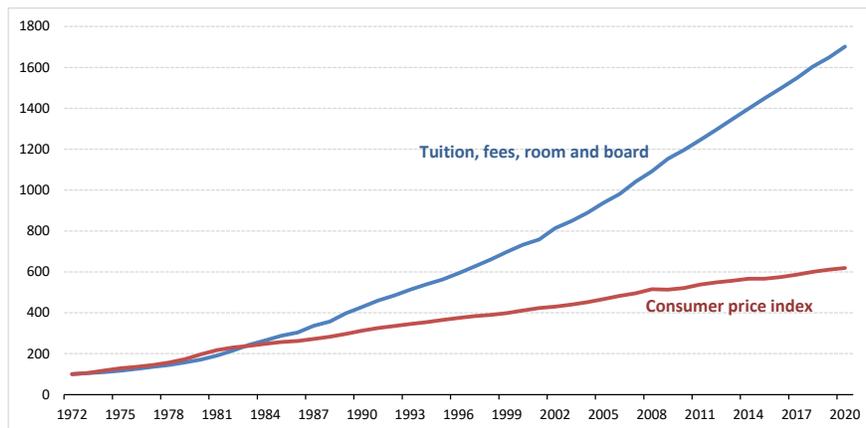

**Figure 1**: Tuition growth has been outpacing inflation. Published tuition, fees, room, and board for private nonprofit four-year institutions compared to the CPI data. For both time series, the base year is 1972. Data source: College Board, Trends in College Pricing 2021, https://research.collegeboard.org/media/xlsx/trends-college-pricing-excel-data-2021-0.xlsx

Several explanations have been offered for increasing tuition. According to Baumol's "cost disease" (Baumol and Bowen, 1965), higher education belongs to a group of service industries that employ highly skilled workers and must pay ever-increasing wages, even though the productivity of labor in those service industries is not increasing fast enough to justify such raises. In higher education, labor productivity has not changed significantly over the past century because of the personalized nature of the services and the perception that improving productivity, for example, by increasing the class size, would degrade the quality of education (Ehrenberg, 2000). However, universities must increase salaries to stay competitive in the labor market against other industries that experience fast productivity growth. The legal profession, healthcare, and the arts have the same cost problem.

Another explanation is Bowen's "revenue theory of cost" (Bowen, 1980), which can be summarized as the five "laws" of higher education costs. First, the main goals of higher education institutions are academic reputation, prestige, and educational excellence.  Second, institutions are willing to spend endless amounts of money in pursuit of those goals. Third, no educational institution ever admits that it has enough money, and therefore, colleges are always looking for more revenue. Fourth, as nonprofits, higher education institutions spend all their revenue. Fifth, the above four laws lead to increases in institutional expenditures.

Then, there is the Bennett hypothesis named after a former Secretary of Education William Bennett. In an op-ed in The New York Times in 1987, he argued that the generous federal student aid policies encourage academic institutions to raise tuition (Archibald and Feldman, 2011: 202). Many studies have investigated this effect with mixed results. For example, Singell and Stone (2007) found evidence that increases in federal financial aid are matched by the tuition growth at private universities as well as non-resident tuition at public universities. However, they found no evidence to support the Bennett hypothesis for the in-state tuition at public universities. Kelchen (2017) found only weak empirical support for the Bennett hypothesis for law schools.

Several studies (Birnbaum, 1988; Frank, 2004; Frank and Cook, 1995; MacLeod and Urquiola, 2021; Winston, 1997; Winston, 2003) have pointed out the strong feedback effects in higher education that reward success with more success, and therefore trigger positional competition between universities for reputation, prestige and resources. If a college improves its reputation and rankings, it gets access to better students, stronger faculty and more external funding (Winston, 2003). Because many more students apply than universities want to admit, top universities can choose the best students (Winston, 1997). But better students, stronger faculty, and the ability to spend more lead to even better reputation, which, again, attracts even stronger students, faculty and more resources (Winston, 2003).  This competitive process leads to a highly differentiated set of colleges (MacLeod and Urquiola, 2021; Winston, 1997).

Numerous other contributions to costs have been identified (see, for example, Ehrenberg, 2000: 14-16; McClure, 2019). They include shared faculty governance, lack of coordination between universities due to the regulatory frameworks, and the organizational structure of universities that encourages inefficiencies. Then, it was also suggested that the prestigious research funding programs encourage exaggerated attention to research by universities. Additionally, college rankings and the amenities competition increase the costs. The following sections discuss college rankings and the facilities "arms race" because they repeatedly come up as reasons behind tuition escalation.

## 3.3   College rankings
College rankings have become a big part of the competitive landscape in higher education. Because educational quality is difficult to determine objectively, the outside world relies on rankings that appear to be scientifically unbiased (Ehrenberg, 2002; Kim, 2018). Rankings have significant sway over students' perception of the quality of education and their decision of where to apply (Spake et al., 2010; Williams, 2022). Besides prospective students, rankings also influence other stakeholders, such as the government, funding agencies, benefactors, faculty, parents, employers, and public opinion (Hazelkorn, 2007; Kim, 2018).

There are numerous rankings of colleges and universities (Hazelkorn, 2007; Miller, 2007: 13). Popular examples include the U.S. News and World Report (USNWR), Forbes, The Princeton Review, The Academic Ranking of World Universities (ARWU)[1] published by Shanghai Jiao Tong University, The Times Higher Education (THE) World University Rankings of more than 1,600 universities worldwide, and The QS World University Rankings[2] of about 1,500 institutions globally.

A typical ranking is based on several criteria that are assigned different weights. For example, The QS World University Rankings rates universities based on six indicators: academic reputation, employer reputation, faculty/student ratio, citations per faculty, international student ratio, and international faculty ratio. Rankings may adjust the criteria and weights from year to year that can shift positions of institutions without any perceptive change in operations or conditions (Ehrenberg, 2002). For example, The California Institute of Technology ranked ninth in 1998 and first in 1999 due to a USNWR ranking methodology change (Monks and Ehrenberg, 1999b). To improve their rankings, universities change their strategy, organization, management, and academics (Ehrenberg, 2002; Hazelkorn, 2007).

### 3.4  Facilities "arms race"

When selecting a college, facilities and non-academic amenities are among the most important factors to students besides educational quality and financial aid (Spake et al., 2010; Goebel et al., 2022). Therefore, to elevate their competitiveness, higher education institutions invest in facilities they hope can attract more students and improve performance (Townsley, 2002; Biemiller, 2015; Wong, 2019). When adding new buildings, universities hope for more tuition revenue and research funding (Harvey, 2020).

Research has observed that a surge in facilities construction due to competition for students and growing enrollments often continues even as enrollments decline (Harvey, 2020; Marcus, 2017). Universities go into significant debt in order to invest in new buildings as a strategy to boost enrollments (Marcus, 2017). Because new buildings add to the annual operating, programmatic, and capital costs that are felt for many years, new facilities may result in the overall negative return, and thus, the facilities "arms race" strategy can be counter-productive in the long run (Harvey, 2020).

---

[1] http://www.shanghairanking.com/ (Last accessed 6/26/22)
[2] https://www.topuniversities.com/qs-world-university-rankings (Last accessed 6/26/22)

# 4 Feedback theory of college competition

Each individual explanation discussed above adds to our understanding of the reasons behind tuition escalation. However, a common theme that runs through all of them is that colleges are participants in a competitive academic market in which they are fighting, without exaggeration, for their survival. According to Moody's, about 11 higher education institutions are forced to close every year due to their financial troubles (Marcus, 2019). The general expectation is that competition in higher education will continue to intensify as the number of college-aged young people in the U.S. declines (Grawe, 2018) and as different countries advance in their efforts to establish world-class universities (Salmi, 2009).

Figure 2 captures the feedback effects of academic competition identified in several earlier studies (Birnbaum, 1988; Frank, 2004; Frank and Cook, 1995; MacLeod and Urquiola, 2021; Winston, 1997; Winston, 2003). Imagine that two peer institutions, College A and College B, compete in the same academic market. When *College A's reputation* improves, College A finds itself in a more advantageous position relative to College B, which leads to more resources being directed to College A. This causal link is shown as two positive arrows between *College A's reputation* and *Resources to College A*. Having access to more resources improves *College A's results* (such as quantity and quality of student applications, more publications by the faculty in more prestigious journals, mentions in the media, etc.) that in turn contribute positively to the *College A's reputation*. This circular causation forms a self-reinforcing loop R1, "winner takes all." On the other hand, as *College A's reputation* improves, it becomes more difficult for College B to attract resources (i.e., recruit students, hire exceptional faculty, and solicit external funding). Lack of resources makes it difficult for College B to produce strong results, which would negatively affect its reputation that would further undermine its position relative to College A. The cycle R2, "no one likes a loser," is a vicious self-reinforcing cycle of College B. The combined forces of reinforcing loops R1 and R2 result in the success-to-the-successful dynamic within the higher education sector.

Assuming College A has a stronger reputation than College B, College B can attempt to improve its situation by initiating an activity that produces results that add to *College B's reputation*. As reputation improves, College B gains ground against College A. This circular causation is the balancing loop B2, "excellence is our goal." If the Advantage of College A over College B is strong, then there is no need for College A to do anything special to strengthen its reputation – this is the negative link in the balancing loop B1, "we are the best." Then, the balancing loop B1 is weak, but College A continues to strengthen its position via the self-reinforcing loop R1. However, if College B succeeds at weakening the positional *Advantage of College A over College B*, College A will be compelled to act to achieve some additional results that would restore its reputation, thus activating loop B1. Feedback loops B1 and B2 describe the mechanism of the positional competition between colleges.

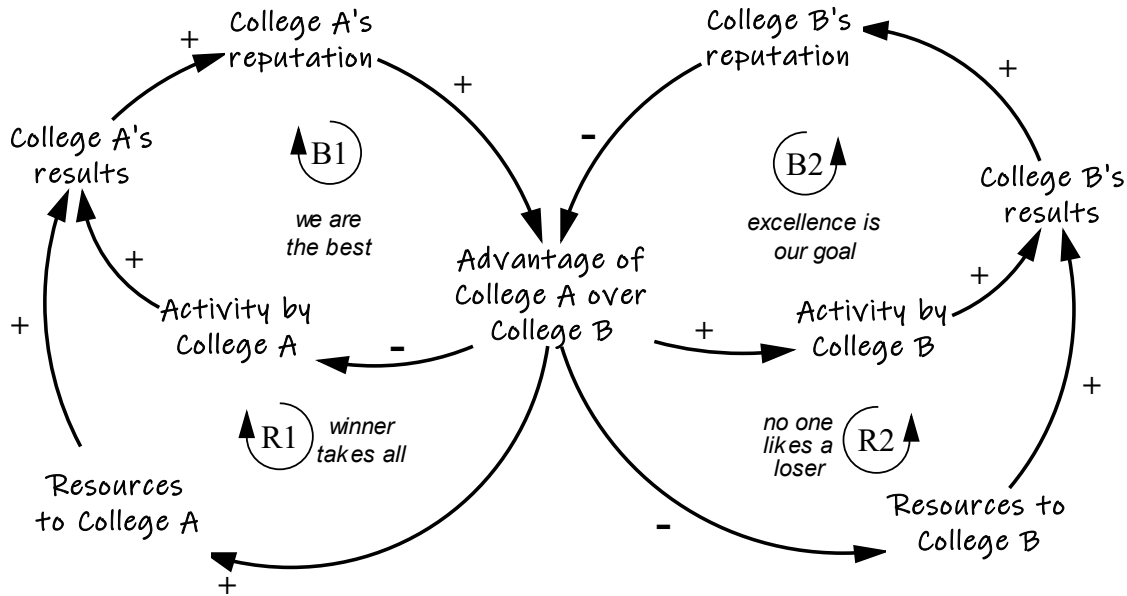

**Figure 2**: The four major feedback loops in the higher education sector. The reinforcing loops R1 and R2 result in the success-to-the-successful dynamics. The positional competition between academic institutions is described by the balancing loops B1 and B2.

The two colleges are subjected to the simultaneous forces of the success-to-the-successful structure of loops R1 and R2, and the positional competition structure consisting of loops B1 and B2. While the market forces of the reinforcing loops R1 and R2 redirect more and more resources to the successful college, the losing college would attempt to weaken the negative effect of the loop R2 by acting through the balancing loop B2. If College B succeeds in improving its reputation relative to College A's reputation, then it will be the turn for College A to take some action to boost its reputation. The hypothesis of this article is that the feedback structure in Figure 2 that describes competition between colleges for reputation and resources would lead to escalating tuition costs.

## 5 The model

To test whether competition for reputation and resources in higher education can lead to escalating tuition costs, we built a formal model that replicates the feedback structures in Figure 2. Figure 3 shows the main components of the model, set as a duopoly consisting of two colleges that vie for reputation, applicants, and tuition money. In our earlier work, we developed a model of a representative private college in consultation with campus stakeholders (Zaini et al., 2017) that was later extended (Pavlov and Katsamakas, 2020; Pavlov and Katsamakas, 2021) in accordance with the economic theory of academic institutions (e.g., Hopkins and Massy, 1981; Massy, 2020). In this article, we use the college model to build a duopoly model that consists of two teaching-focused colleges that vie for reputation, applicants, and the tuition money. To keep things simple, this model assumes that colleges compete only on facilities. College applicants choose between the two colleges based on the rankings and the financial aid offered by the colleges. An independent ranking agency determines the relative order of the colleges -- each college is ranked as #1 or #2. The college rank information is available to potential students. More students apply to the top-ranked college than the second-ranked college.

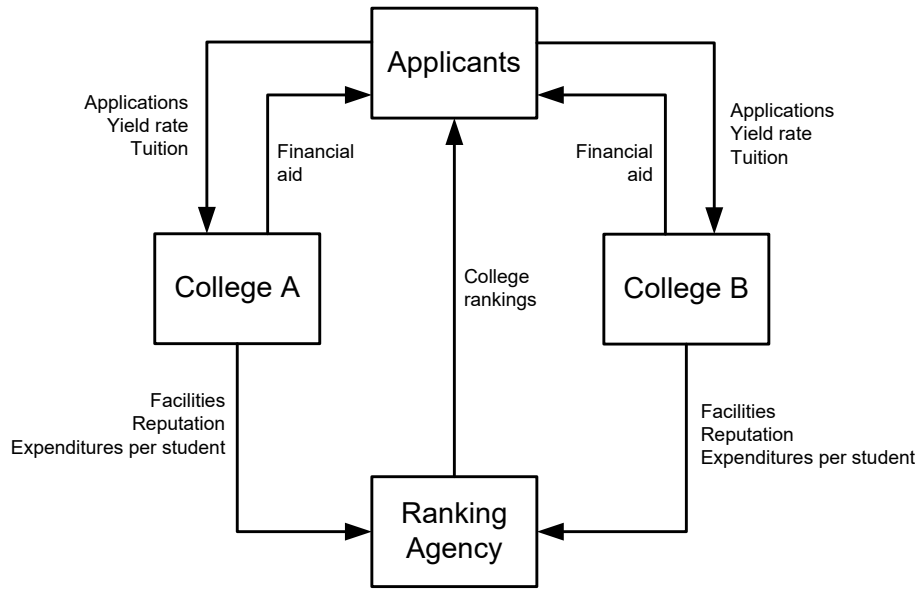

**Figure 3**: The model consists of two competing colleges, a pool of applicants, and a ranking agency.

Our model includes the following assumptions:
- Each college tries to improve its ranking.
- Colleges compete by improving facilities.
- Colleges are similar, but not necessarily identical in terms of their parameter values and strategies they choose.
- There is no shortage of applicants to colleges.
- There is no limit on borrowing by students.
- Colleges have no problem attracting qualified faculty.
- Colleges are teaching-focused.

Additionally, our model differs from the competition model in Epple et al. (2006) in several ways. While Epple et al. (2006) focus on equilibrium and treat tuition and endowments as exogenous, our interest in tuition inflation demands the out-of-equilibrium analysis, and as a result, our model incorporates tuition adjustments and allows for endowment growth. Since our objective is not to explain college stratification, our model does not differentiate students by household income and academic ability.

Below we provide the mathematical description of the model. The list of all variables and parameter values as well as their units can be found in the Appendix.

## 5.1 Colleges

The college agent describes operations of a stylized private tuition-dependent non-profit college. Each college agent is organized as four sectors: students, faculty, facilities, and financials. The sectors are described below.

### 5.1.1 Students

The number of students changes according to the equation:

$$\frac{dY}{dt} = i_Y - o_Y,$$

where incoming class, $i_Y$, adds to the total enrollments and then students graduate at the rate $o_Y = Y/\tau_Y$, where $\tau_Y$ is the average time to graduation. The incoming class is equal to the yield, $i_Y = yA_Y$, where $0 < y < 1$ is the yield rate and $A_Y$ are admitted applicants. The yield is less than the number of admitted students, $i_Y < A_Y$, because some admitted students do not matriculate. The admit rate is the ratio of admitted applicants to the number of all applicants to the college, $\lambda_Y = A_Y/A$. Each college keeps track of its typical admit rate, $\bar{\lambda}_Y$, and its typical yield rate, $\bar{y}$. We assume that each college tries to enroll an incoming class of size, $\bar{Y}$. Therefore, given the typical admit rate, $\bar{\lambda}_Y$, and the typical yield rate, $\bar{y}$, the desired number of applications for a college that would result in the incoming class, $\bar{Y}$, is $A^* = \bar{Y}/(\bar{\lambda}_Y \bar{y})$.

The college charges tuition, fees, room, and board price $P$ (henceforth, we will simply call it tuition). Then, by definition, the average net tuition, $NP$, is the sticker price, $P$, minus the average financial aid per student, $\bar{\varphi}$, that is $NP = P - \bar{\varphi}$.

The yield rate is $y = \bar{y}f_y(NP)$, where $\bar{y}$ is the typical yield rate for a college that is adjusted for financial aid by the function $f_y(NP)$ that captures the effect of the financial aid on yield. Here $f_y(NP) = (\overline{NP}/NP)^\alpha$, where $NP$ is the average net tuition, $\overline{NP}$ is the typical net tuition, and $\alpha$ is the net tuition elasticity of yield. We assume that the typical net tuition, $\overline{NP}$, is an exponential smooth of the average net tuition for the past several periods of the simulation.

The college adjusts its financial aid per student, $\varphi$, in response to the level of applications: $\varphi = \bar{\varphi} A^*/A^i$, where $\bar{\varphi}$ is the typical financial aid per student offered by the college over the past several simulated years, $A^*$ is the desired number of applications, and $A^i$ is the number of applications that college $i$ would expect to receive without adjusting the financial aid.

Student satisfaction, $Q_Y$, is a function of faculty academic experience, $Q_F$:

$$Q_Y = g_Y(Q_F), \ g'_Y(Q_F) > 0$$

See Figure 13 in the Appendix for definition of the graphical function $g_Y(Q_F)$. The faculty academic experience, $Q_F$, is defined below in the *Faculty* sub-section.

To measure college prestige, we follow Caskey (2018) by introducing an Index of Selectivity, $Q_A$, which is a weighted average of the rejection rate, $1 - \lambda_Y$, and yield rate, $y$:

$$Q_A = \frac{(1 - \lambda_Y) + y}{2}$$

College reputation, $Q$, is the weighted average of student satisfaction, $Q_Y$, and Index of Selectivity, $Q_A$:

$$Q = \frac{Q_Y + Q_A}{2}$$

### 5.1.2 Faculty

The average academic load that an instructor teaches per year is $L_Y = Y l_Y / F$, where $l_Y$ is the number of academic credits that a typical student takes in a year, $Y$ is the number of enrolled students and $F$ is the number of employed instructors. If we introduce the student-to-faculty ratio $q_{YF} = Y/F$, then the above equation can be rewritten as

$$L_Y = q_{YF} l_Y \ .$$

Given the standard student to faculty ratio, $\bar{q}_{YF}$, the standard faculty load is

$$\bar{L}_Y = \bar{q}_{YF} l_Y$$

The faculty academic load index is $l_F = L_Y / \bar{L}_Y$.

The number of faculty, $F$, employed by the college increases when new professors are hired, $i_F$, and it decreases due to faculty attrition, $o_F$:

$$\frac{dF}{dt} = i_F - o_F$$

Faculty hiring depends on whether there is a faculty shortage, which is measured as a fraction $\nu = max(0, l_F - 1)$, where $l_F > 0$ is the faculty academic load index. The normal teaching load per faculty is $l_F = 1$. When professors teach less than the standard load, $0 < l_F < 1$, and therefore the shortage index is $\nu = 0$. When the teaching load exceeds the normal level, i.e. $l_F > 1$, the faculty shortage is $\nu = l_F - 1 > 0$. When $\nu > 0$, given some hiring delay $\tau_h$, more faculty will be hired:

$$i_F = \frac{\nu F}{\tau_h}$$

The faculty academic experience, $0 < Q_F \leq 1$, is an index that depends on the faculty space loading, $\gamma_B$ (defined in the Facilities sub-section), and the academic teaching load, $l_F$: $Q_F = g_F(x)$, $g'_F(\cdot) < 0$. Here $x = 0.5l_F + 0.5\gamma_B$ is a composite variable consisting of faculty academic load index $l_F$ and faculty space loading $\gamma_B$. As teaching load and faculty space loading increase, the faculty academic experience declines. See Figure 14 in the Appendix for definition of the graphical function $g_F(x)$.

The faculty attrition is:

$$o_F = (\phi/Q_F)F ,$$

where $\phi/Q_F$ is the annual faculty turnover rate that depends on the typical turnover rate $0 < \phi < 1$ and the faculty academic experience, $Q_F$. If the faculty academic experience deteriorates, the faculty morale declines and faculty attrition increases.

### 5.1.3 Facilities

Facilities include office space for the faculty and student facilities comprising classrooms, laboratory space for teaching and dorm rooms. Because capital projects are complex undertakings that involve many stakeholders and take many years of planning, fundraising, and construction (Balderston, 1995: 239; Harvey, 2020), available space often lags behind the desired space. Therefore, we assume that student facilities are represented by two stocks: student facilities that have been constructed, $B_Y$, and student facilities that are planned, $K_Y^i$, but will take some time $\tau_B$ to construct. The construction rate for student facilities is:

$$\frac{dB_Y}{dt} = \frac{K_Y^i}{\tau_B}$$

We assume that colleges compete on student facilities. Each college constantly compares existing and planned student facilities to the facilities at the other college. Then, if college $j$ has $b_Y^j$ student facilities per student, then to be competitive, college $i$ wants to have $(1+k)b_Y^j$, $j \neq i$ student facilities per student, where $0 \leq k \leq 1$ is a competitive edge in facilities. In other words, a college would want a slightly better student facilities than its peers. Then, the student facilities gap that college $i$ wants to fill is $max(0, (1+k)b_Y^j Y^i - (B_Y^i + K_Y^i))$, where the term $B_Y^i + K_Y^i$ is the stock of existing and planned student facilities at college $i$.

Assuming that each faculty member requires some office space, $b_F$, the total faculty space needed is $\bar{B}_F = b_F F$. However, due to construction delays, the available faculty space, $B_F$, might be different than the space needed. This potential discrepancy is measured by the faculty space loading ratio that is defined as $\gamma_B = \bar{B}_F/B_F$.

### 5.1.4 Financials

Expenditures include faculty compensation, $C_F$, operating and maintenance cost of facilities, $C_B$, debt payments, $C_D$, and the financial aid expenditures, $C_A$:

$$C = C_F + C_B + C_D + C_A$$

Faculty compensation is $C_F = \omega F$, where $\omega$ is the average compensation of a faculty member and $F$ is the number of faculty employed by the college. The operating and maintenance cost of facilities is $C_B = c_B B$, where $c_B$ is the operating cost of facilities and $B$ is the total facilities. The financial aid expenditures term is the product of the average financial aid per student and the stock of enrolled students, $C_A = \bar{\varphi} Y$.

The revenue is the sum:
$$R = R_Y + o_m + o_E + g_u$$

Here, tuition revenue is the product of the sticker price and enrolled students, $R_Y = PY$, the term $o_m$ is the liquidity draw, $o_E$ is the endowment draw, and $g_u$ are the unrestricted gifts that can be used for operations. When the difference between revenue and expenditures, $S = R - C$, is positive it is a surplus, otherwise it is a deficit.

When net revenue is negative, $S < 0$, the deficit per student is $S/Y$. To resolve this deficit, colleges increase tuition. The model assumes that colleges can annually increase tuition by at most $\beta$ percent. Then, if there is a deficit, the college increases its tuition, $P$, by $\min(S/Y, \beta P)$. Thus, the sticker price adjustment process is:

$$\frac{dP}{dt} = \min(S/Y, \beta P)/\tau_P,$$

where $\tau_P$ is a constant denoting the adjustment speed.

The model assumes that the college maintains a stock of liquid assets, such as cash, which is replenished when the college has an operating surplus and depleted when the college needs to use the cash for operations. Maintaining a cash reserve for a "rainy day" is an approach to strengthening the financial health of a college (Townsley, 2009). Liquid assets increase when there is a surplus, $S > 0$, and decreases when cash, $o_m$, is used to pay for operations:

$$\frac{dM}{dt} = S - o_m$$

When there is a deficit, that is $S < 0$, to cover the deficit, a college increases tuition by the amount of $|S|/Y$.

Most of the capital funding comes in the form of loans (Balderston, 1995: 239); therefore, this model assumes that facilities are constructed with borrowed funds. Debt, $D$, changes as

$$\frac{dD}{dt} = d_B + m_D - p_D$$

where $d_B$ are the funds borrowed for new construction, money borrowed for operations is $m_D$, and $p_D$ are payments of the debt principal. Debt payments, $C_D$, are principal payments, $p_D$, plus annual accrued interest payments, $r_D D$, where $r_D$ is the debt interest rate:

$$C_D = p_D + r_D D$$

Colleges strive to increase their endowments. An endowment increases due to the good performance of the college's investments in the stock market or through the gifts. The endowment adjusts according to the equation:

$$\frac{dE}{dt} = g_r + r_E E - o_E .$$

Here $g_r$ are the gifts and $r_E E$ is the term describing investment earning on the endowment when the investment rate of return is $r_E$. Endowment withdrawals are $o_E$.

## 5.2 The ranking agency

The model includes a stylized ranking agency that compares the two colleges. The ranking agency receives data from universities and then communicates college rankings to the applicants.

The ranking agency calculates a ranking index, $I$, for each college. The index for College A is a weighted relative position of College A with respect to College B in terms of its reputation, $Q^A$, expenditures per student, $e^A$, and student building space per student, $b_Y^A$.

$$I^A = \alpha_R \frac{Q^A}{Q^A + Q^B} + \alpha_e \frac{e^A}{e^A + e^B} + \alpha_f \frac{b_Y^A}{b_Y^A + b_Y^B}$$

In this equation, $\alpha_R, \alpha_e,$ and $\alpha_f$ are the weights. Then each college, for example College A, is assigned a rank according to the following rule:

$$Rank^A = \begin{cases} 1 & I^A \geq I^B \\ 2 & I^A < I^B \end{cases}$$

## 5.3 Applicants

Schools that manage to improve their rankings see more applicants (Monks and Ehrenberg, 1999b). Therefore, in this model, we assume that College A receives $A^A$ applications from the application pool $A$ that depends on the respective ranks of College A and College B:

$$A^A = A\left(1 - \frac{Rank^A}{Rank^A + Rank^B}\right)$$

In this formulation, when two colleges are of equal reputation, they will receive an equal number of applications. However, if a college ranks as number one, then it receives 2/3 of the applicant pool $A$ and the other college receives the remaining 1/3 of the applications.

# 6 Scenarios

The model was implemented and simulated in the *Stella Architect* software from *isee systems*. Below we show results for five scenarios. Scenario S1 is the base run that introduces competition. It demonstrates that contrary to the conventional wisdom that competition lowers prices, positional competition between academic institutions leads to escalating college tuition costs. In Scenarios S2-S5, the strategy of College A remains unchanged; however, College B attempts various policies. Scenario S2 confirms an observation (2004) that a college cannot drop out of a positional arms race without severe consequences. Scenario S3 demonstrates that a capital gift can help a college gain an edge over its competition as the gift strengthens the balancing loop B2 in Figure 2. Scenario S4 simulates a case when a college strategically strengthens loop B2. Scenario S5 combines an aggressive facilities build-up with a capital campaign, which intensifies the competition.

## 6.1 S1: Rankings are introduced

The simulation starts in the steady state in Year 1. We assume that at the beginning, colleges are not engaged in positional competition. However, in Year 5, rankings are introduced, and colleges begin to compete. Slight variations in initial variable values ensure that College A has a slightly better reputation at the start of the simulation that positions it as #1 in the rankings, while College B is ranked #2 (Figure 4a). College B responds by building up facilities (Figure 4b) that ensure that by Year 9 it takes the first spot in the rankings (Figure 4a). As soon as College A loses its dominant position in the rankings, College A responds by improving student facilities (see blue line in Figure 4b). This positional competition is described by loops B1 and B2 in Figure 2.

A top-ranked college is rewarded with more applications (Figure 4c). For example, applications to College A surge in Year 5. However, as College B acts through loop B2 (Figure 2) by building up facilities and earning the first spot in the rankings, College B eventually attracts more applications than College A (Figure 4c). Note that initially colleges earn enough tuition revenue, so they run surpluses, and therefore there is no need to increase tuition until about Year 20 (Figure 4d). However, after Year 20, the competition leads to escalating tuition. During the simulation, College A is ranked first 52 percent of the time, while College B is ranked first during 48 percent of the time.

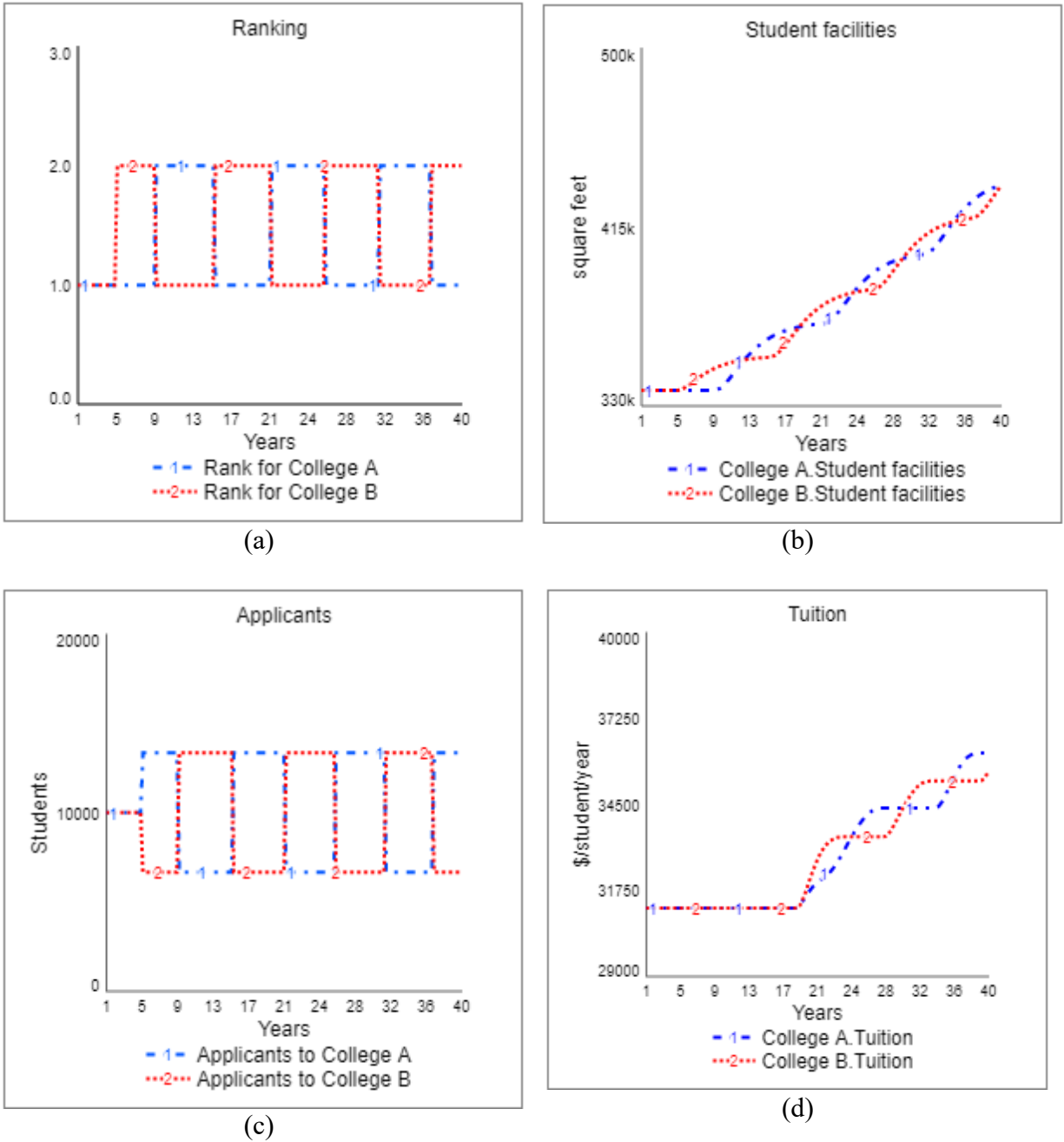

**Figure 4**: Positional competition leads to a facilities buildup and escalating tuition

Monks and Ehrenberg (1999a) observed that when institutions drop in the ranking order, they tend to give more financial aid in order to attract enough students. Figure 5 shows the same simulated pattern. As an example, we picked College B. Whenever College B switches to the second position in the rankings (see the red dotted line), it has to offer more financial aid to fill the incoming class, and that pushes up the discount rate. The discount rate is the fraction of the tuition revenue that is given as the financial aid. The fact that it is easier to attract students for prestigious universities is a feature of the winner-take-all educational markets. The feedback loops that are responsible for this behavior are R1 and R2 in Figure 2.

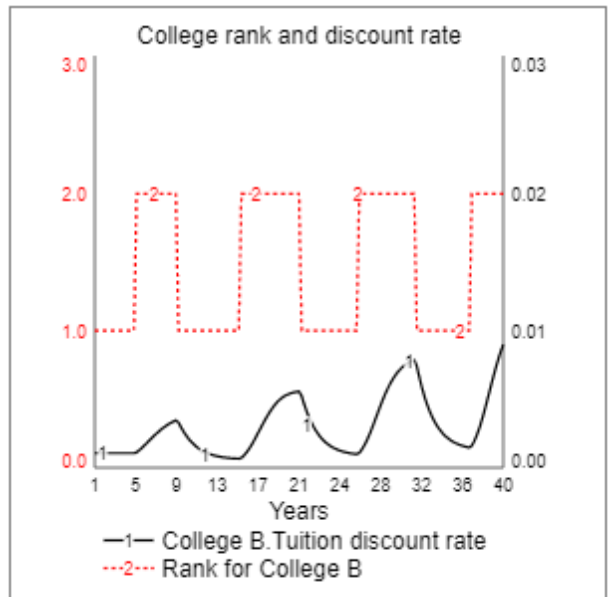

**Figure 5**: When a college drops in the ranking order, it must offer more financial aid that increases the discount rate. The shown trajectories are for College B. College A exhibits the same behavior.

## 6.2   S2: One college ignores rankings

This scenario tests the possibility that a college drops out of the competition. We assume that College B makes a bold move and stops paying attention to the rankings in Year 25. In effect, College B eliminates the feedback loop B2. College B no longer tries to match the facilities of College A (Figure 6a). Notice that for a while, College A continues adding facilities that had already been planned (blue curve in Figure 6a). As expected, College B slips to second place in the rankings (Figure 6b). As a result, College B becomes less selective – the admit rate jumps from less than 20 percent when it was ranked number one to nearly 40 percent (Figure 6c). However, eliminating competition stops tuition growth (Figure 6d). As might be expected, College A dominates the rankings by being in the first spot during 67 percent of the time.

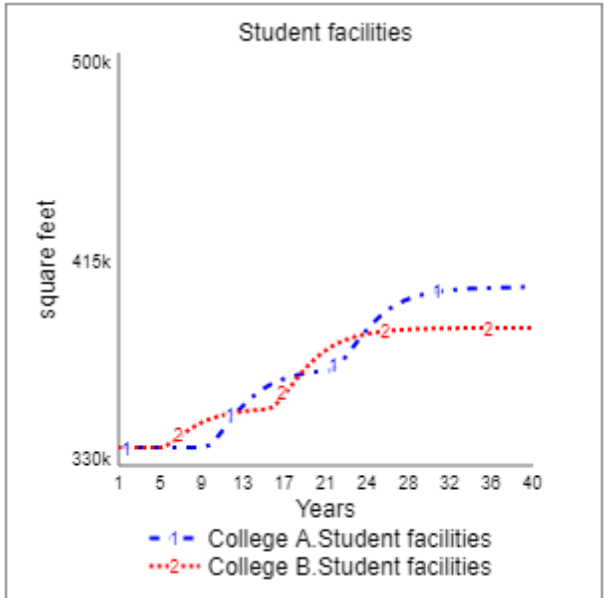 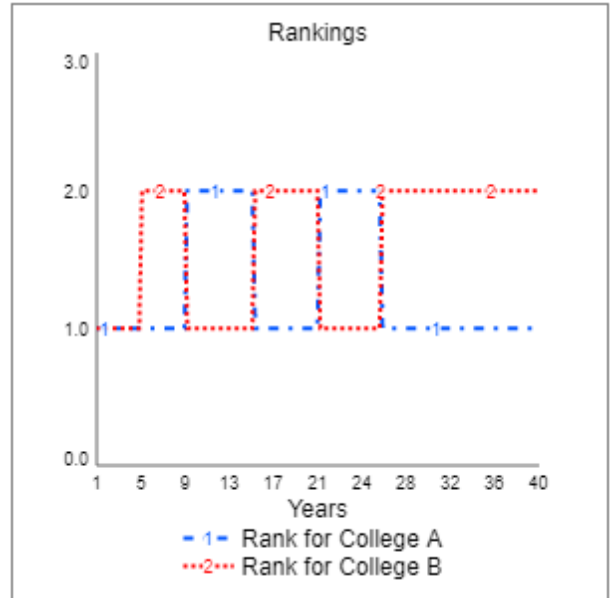
(a) (b)

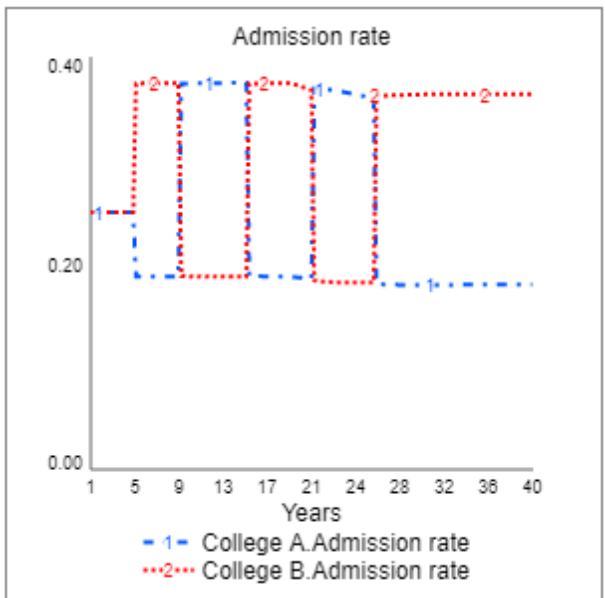 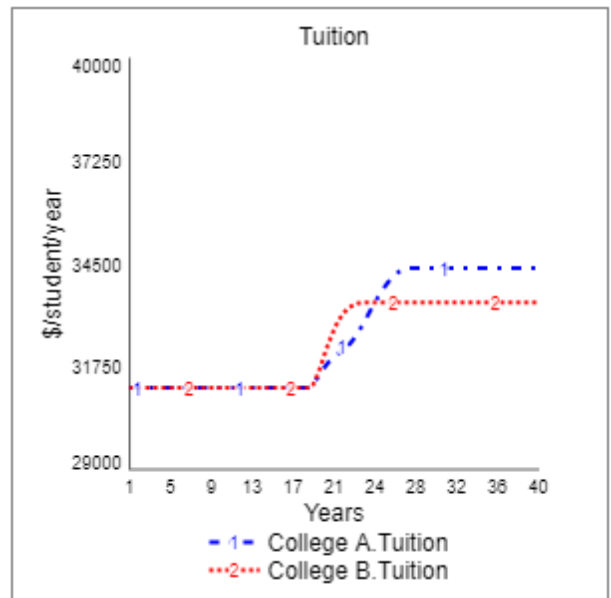
(c ) (d)

**Figure 6**: College B ignores rankings starting in Year 25. As a result, College B stays in the second position in the rankings, and it receives fewer applications than College A. But eliminating competition stops tuition escalation.

## 6.3 S3: A capital campaign

This scenario assumes that College B raises external funds for new construction through a capital campaign. College B receives a large one-time donation in Year 20 that reduces the need to borrow for capital projects.

The capital campaign does not eliminate any of the loops in Figure 2. Therefore, since the feedback structure does not change, the competition through student facilities continues (see Figure 7a) as in the earlier experiments, and tuition still increases (Figure 7b). The rankings exhibit the same alternating pattern (Figure 7c) as before. However, the gift allowed College B to borrow less for construction, which put it in a stronger financial position than College A (see Figure 7d).

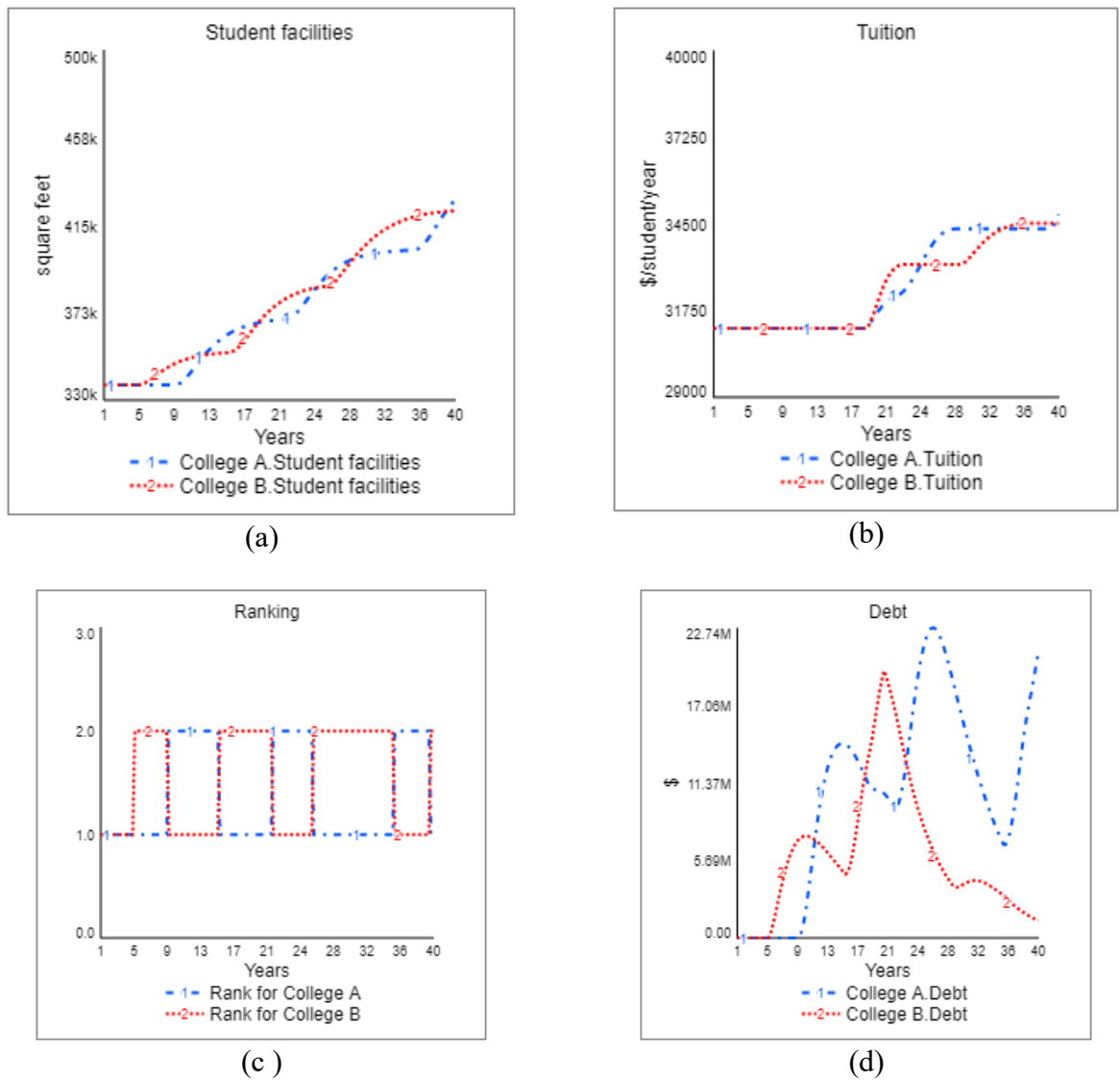

**Figure 7**: A capital campaign by College B improves its financial position but does not eliminate competition; therefore, it does not stop tuition escalation.

## 6.4 S4: An aggressive facilities buildup

Throughout previous simulations we assumed that both colleges compete by building facilities that exceed their competition by five percent, i.e., $k = 0.05$. In this scenario, College B undertakes an aggressive facilities build-up starting in Year 25. The college decides that it needs facilities that exceed its competition by 20 percent. While this decision strengthens loop B2 in Figure 2, it does not modify the feedback structure. All loops remain; therefore, the feedback theory predicts, and the simulation confirms, that the behavior patterns do not change. There is still a facilities buildup (Figure 8a), tuition grows to support greater expenditures (Figure 8b), rankings oscillate (Figure 8c), and colleges accumulate debt which weakens their financial positions (Figure 8d). However, in this scenario, College B ranks higher more frequently than in the base-case Scenario S1. Here, College B occupies the first spot in the rankings in 53 percent of the years, which is more than the 48 percent in Scenario S1.

## 6.5 S5: An excellence campaign

This scenario combines an aggressive facilities buildup with a capital campaign. As in Scenario S3, College B receives a gift in Year 20 for capital projects. This situation is followed by an aggressive facilities buildup starting in Year 25, as in Scenario S4. These two actions strengthen loop B2 in Figure 2 without modifying the overall feedback structure of the competition between colleges. Accordingly, the feedback theory predicts that the system behavior is not likely to change, but we may expect better results for College B.

Simulation results for this scenario are in Figure 9. With more capital funding and a desire to lead in facilities, College B has more student facilities nearly every year after Year 25 (see Figure 9a). Tuition is still increasing, even though the better financial position of College B allows it to underprice College A nearly every year after Year 25 (Figure 9b). Also, College B leads the rankings in 58 percent of the years (Figure 9c), which is better than in Scenarios S3 and S4. This implies that the availability of additional funds enabled College B to be a stronger competitor against College A while incurring less debt (Figure 9d).

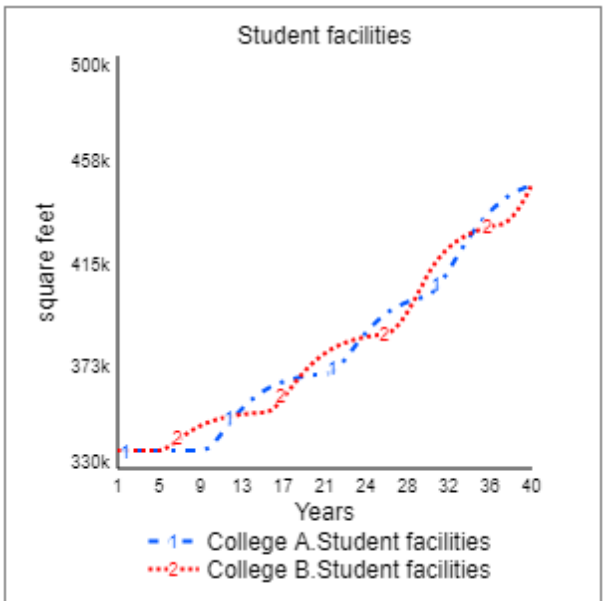

(a)

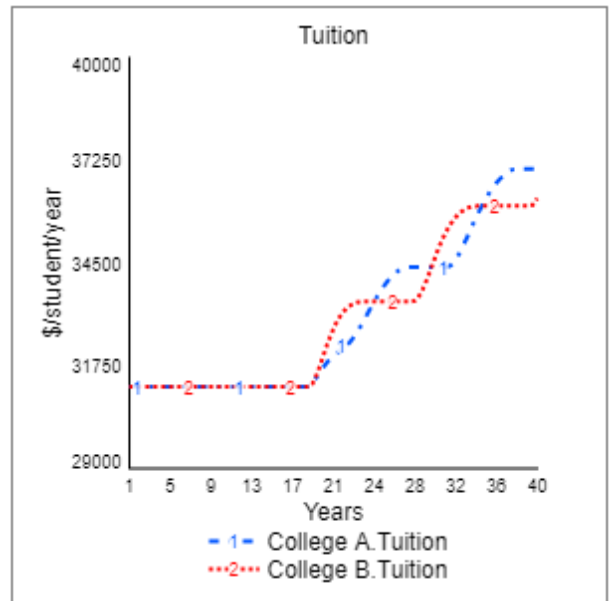

(b)

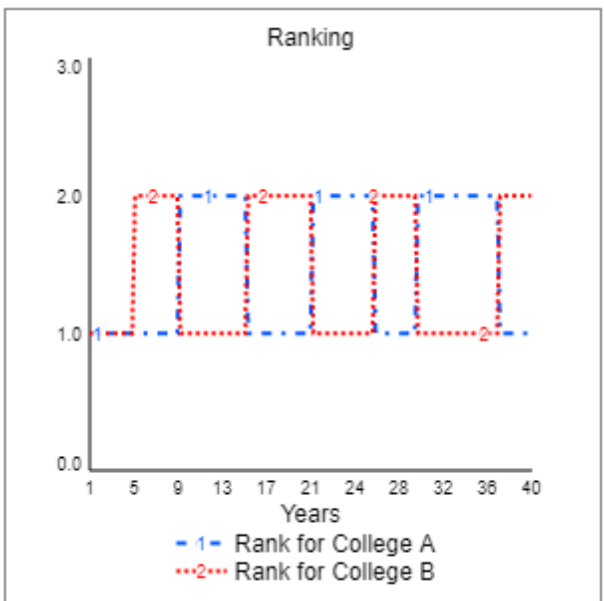

(c)

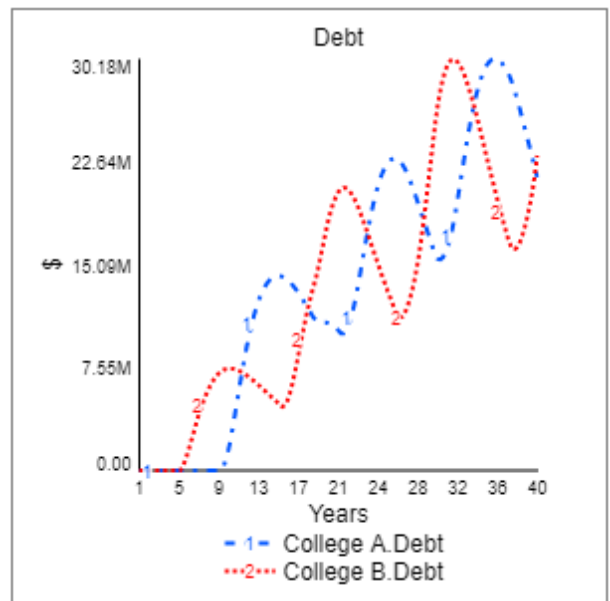

(d)

**Figure 8**: An aggressive buildup by College B starting in Year 25 improves its performance in the rankings but does not change behavior patterns and does not eliminate tuition escalation.

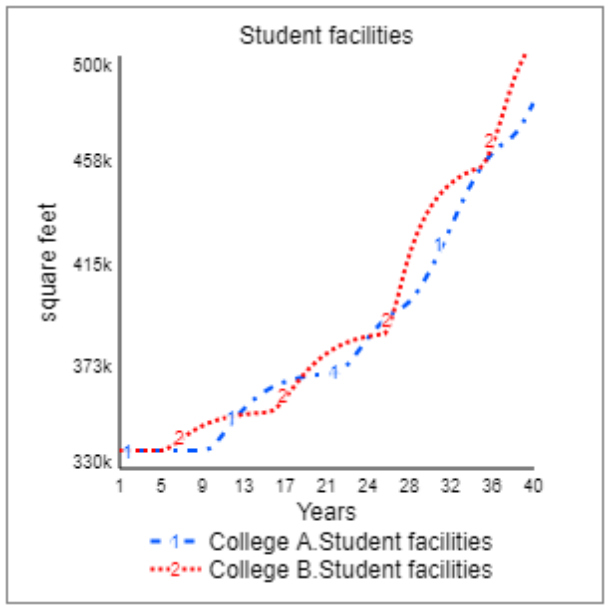
(a)

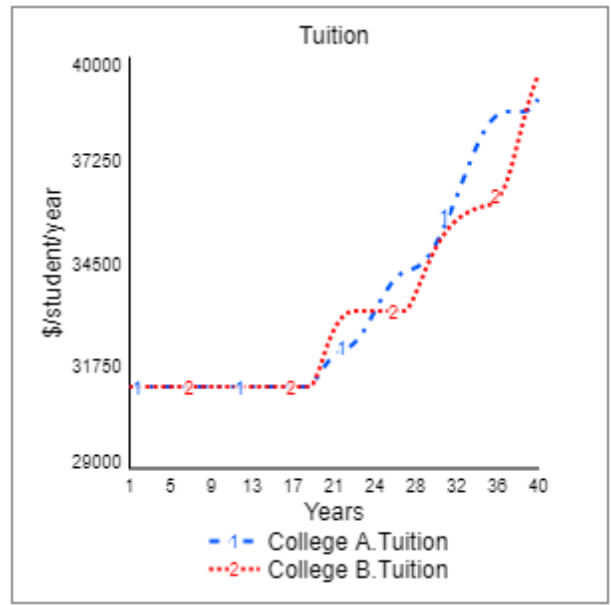
(b)

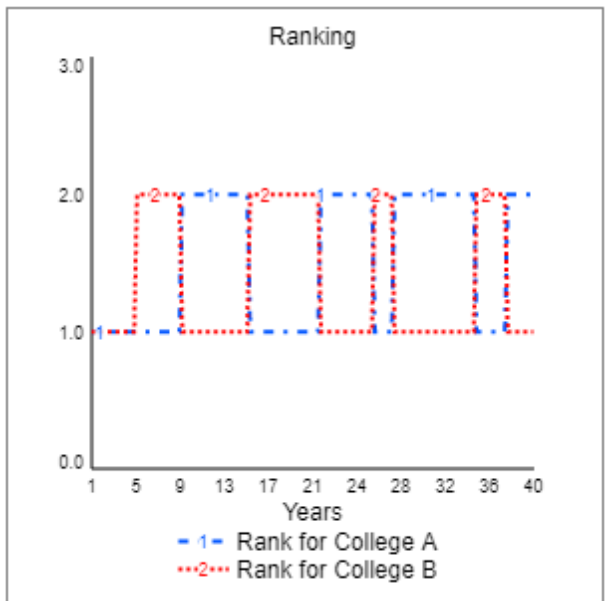
(c )

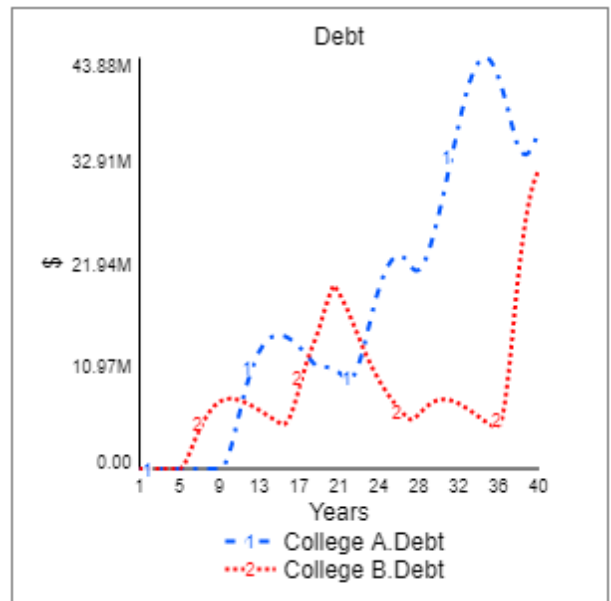
(d)

**Figure 9**: An excellence campaign by College B allows it to dominate the market while maintaining a more sustainable financial position.

## 6.6 Scenario comparison

This section compares the five scenarios by examining the industry-wide patterns, which are shown in Figures 10-12. The trajectories were obtained by averaging tuition, institutional debt, and per student expenditures for College A and College B. As can be seen in Figure 10, the only scenario in which tuition growth stopped was S2, which is when College B quit the positional competition with its peer College A. By disregarding the rankings, College B eliminated balancing loop B2 in Figure 2, thus changing the feedback structure of the market. None of the other scenarios modified the competition feedback structure in Figure 2. The fastest tuition growth occurred in Scenario S5 in which one of the colleges was aggressively pursuing better ranking while also raising external funds through a capital campaign. These two actions by College B in Scenario S5 strengthened loop B2 that intensified the competition between the colleges and resulted in higher costs and higher tuition.

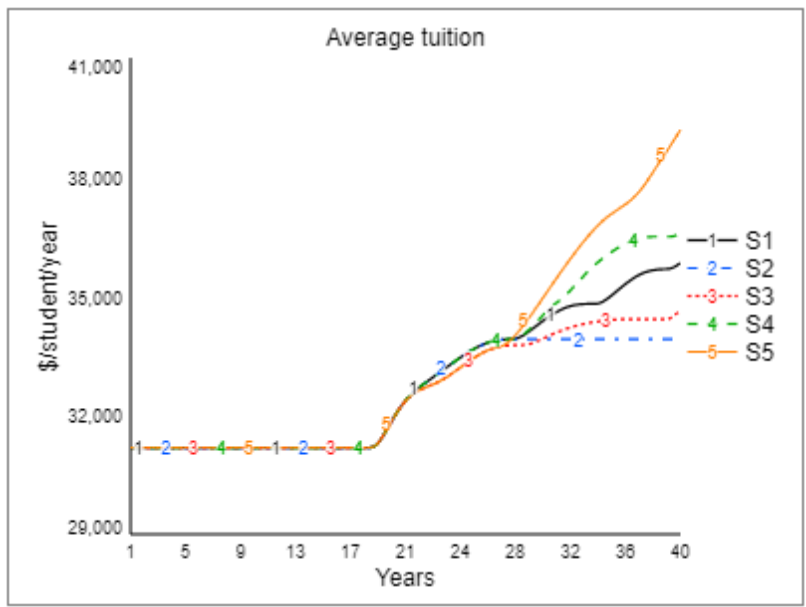

**Figure 10**: Average tuition increases in all the scenarios except S2, when one of the colleges dropped out of the positional competition.

Figure 11 shows a range of outcomes for the average institutional debt. The industry accumulates the highest debt in Scenario S5 when College B aggressively builds up facilities while supporting it through a capital campaign. Scenario S2 has the lowest institutional debt.

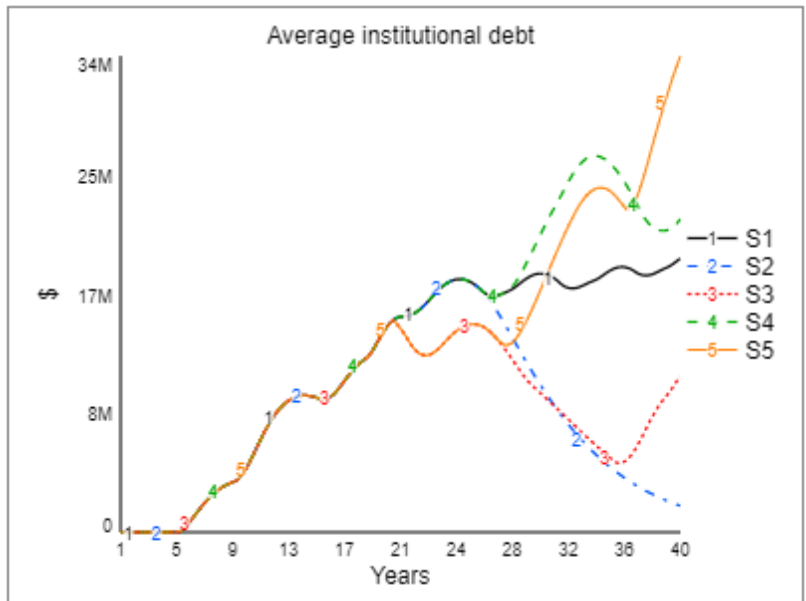

**Figure 11**: Scenarios produce a range of outcomes with respect to average institutional debt.

The patterns in Figure 12 show a range of possibilities for the average expenditures per student. Education provision is the least expensive in Scenario S2 and the most expensive in Scenario S5.

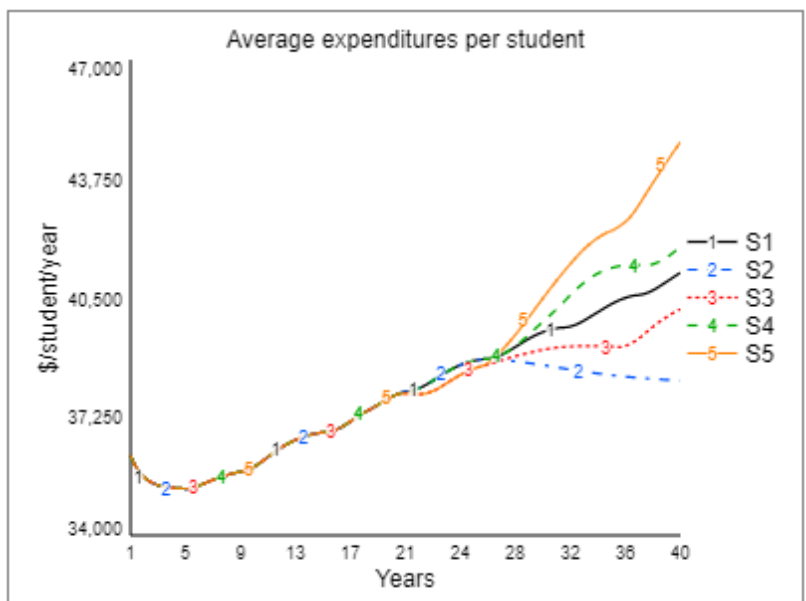

**Figure 12**: The average expenditure per student for the industry during the five scenarios.

Taken together, these simulations demonstrate that competition between academic institutions for resources and reputation leads to tuition escalation that negatively affects students and their families. Four of the five scenarios – rankings, a capital campaign, facilities improvements, and an excellence campaign – increases college tuition, institutional debt, and expenditures per student; only the scenario of ignoring the rankings decreases these measures.

# 7 Discussion

Our computational experiments confirm that the feedback model of college competition can explain tuition inflation. The feedback theory developed in this article suggests that tuition is stable without positional competition between colleges, as is the case at the start of Scenario S1 and later in Scenario S2. However, the college that stopped competing in Scenario S2 slipped in the rankings and received fewer applications. This result agrees with an observation made by Frank (2004) that no single higher education institution can escape on its own a positional arms race without severe consequences. In other words, without structural changes in the higher education market, the positional race will not end until colleges run out of resources.

To have a sustainable higher education sector that is not crippled by escalating costs and tuition, the sector needs new ideas on organizing higher education without encouraging intense competition between academic institutions (Naidoo, 2016). Some possible solutions are:
- Weakening the connection between the college rank and associated rewards would weaken reinforcing loops R1 and R2 which would reduce positional competition.
- Reducing uncertainty of the college quality would diminish the need for positional competition, thus weakening balancing loops B1 and B2 in Figure 2.
- A new ranking category can be introduced for "value colleges" that provide quality education without the amenities of more expensive institutions. In a recent survey, about 40 percent of students indicated that they would be open to quality education with fewer amenities if it was less expensive (Goebel et al., 2022). This approach would not dismantle the feedback structure of academic competition in Figure 2. Instead, it would create a parallel structure for colleges ranked on an alternative set of criteria.
- Rankings can be modified to encourage collaboration (Ehrenberg, 2002). This approach would weaken competition loops in Figure 2 and motivate colleges to seek economies of scale and scope by pooling resources.
- Randomly assigning students to colleges would weaken reinforcing loops R1 and R2 in the academic market.
- To stop the "arms race", The US Department of Justice may reconsider its ruling from 1991 that prohibits coordination between universities (Ehrenberg, 2000: 14; Frank, 2004).

# 8 Conclusion

This article shows that competition between colleges that vie for reputation and resources triggers tuition inflation. The article makes several recommendations on how to control tuition escalation. Besides contributing to the research on the economics of higher education and economics of positional competition, this article also demonstrates how to apply feedback analysis to economic problems, which is the domain of feedback economics (Cavana et al., 2021). We hope that this article will encourage more economists to add feedback analysis to their research toolset.

This work can be extended in several ways. Future models can include the academic quality of student applications and incorporate the cost of student recruitment. The theoretical framework developed in this article can be adapted for research universities by considering the tradeoffs between teaching and research, the ability of universities to attract research faculty, and the inclusion of research funding in the operating budget. We could also explore the effects of competition and various financial aid packages on the diversity of students. Finally, future research can examine if competition and tuition can be reduced by introducing different college rankings schemas.

# 9 Appendix: Variables and parameters

This Appendix shows variables, parameter values, and their units. The model has been checked for unit-consistency. The initial values for state variables are for the steady state that exists when colleges are not engaged in positional competition. The model was implemented and simulated in the *Stella Architect* software available from *isee systems*.

**Students**

| Variable | Notation | Units | Notes |
|---|---|---|---|
| Applications to college $i$ | $A^i$ | students/year | Colleges with better reputation receive significantly more applications than they can admit. |
| Application target | $A^*$ | students/year | College needs to receive this many applications to meet its yield target. |
| Admit target | $A_Y^*$ | students/year | This is how many students a college wants to admit considering that not everyone will matriculate. |
| Admitted students | $A_Y$ | students/year | If applications are low, a college may admit fewer students than the desired admit target. |
| Admit rate | $\lambda_Y$ | dimensionless | Selective colleges have lower admit rates. |
| Typica admit rate | $\bar{\lambda}_Y$ | dimensionless | The average value of the admit rate over several years. |
| Yield | $i_Y$ | students/year | Yield is the incoming class. |
| Yield rate | $y$ | dimensionless | A fraction of admitted students who enroll. |
| Net tuition elasticity of yield | $\alpha = -0.3$ | dimensionless | Denning (2017) estimate that price elasticity of enrollment for colleges in Texas is -0.29. We assume a value within that range. |
| Student satisfaction | $Q_Y = g_Y(Q_F)$ | dimensionless | Student satisfaction is a function of faculty academic experience. College reputation depends on student satisfaction. See Figure 13 for definition of $g_Y(Q_F)$. |

| Variable | Notation | Units | Notes |
|---|---|---|---|
| Index of selectivity for college $i$ | $Q_A^i$ | dimensionless | Measures college prestige. An average of the rejection rate, $1 - \lambda_Y$, and yield rate, $y$. |
| College reputation | $Q$ | dimensionless | This is an index. Assumed initial value is $Q_0 = 1$. |
| Graduates per year | $o_Y = Y/\tau_Y$ | students/year | $\tau_Y = 4.5$ years is the average time to graduation |
| Students | $Y$ | students | Initial value is $Y_0 = 3{,}375$ students. |
| Target annual enrollment | $\bar{Y} = 750$ | Students/year | Assumed value based on discussions with stakeholders for Zaini et al. (2017). |

**Faculty**

| Variable | Notation | Units | Notes |
|---|---|---|---|
| Student to faculty ratio | $q_{YF} = Y/F$ | students/faculty | More selective institutions have lower student to faculty ratios. |
| Standard student to faculty ratio | $\bar{q}_{YF} = 5$ | students/faculty | From Pavlov and Katsamakas (2020) |
| Average generated load per student | $l_Y = 100$ | credits | A measure of how many courses a student takes per year. From Pavlov and Katsamakas (2020) |
| Average faculty academic load | $L_Y$ | credits/faculty | Average teaching load |
| Standard faculty load | $\bar{L}_Y$ | credits/faculty | |
| Faculty academic load index | $l_F$ | dimensionless | Faculty academic experience declines when their academic load is above the standard level. |
| Faculty academic experience | $Q_F = g_F(l_F, \gamma_B)$ | dimensionless | See Fig 14 for definition of $g_F(\cdot)$ |
| New faculty | $i_F$ | faculty/year | An inflow into the stock of faculty |
| Time to decide to leave | $\tau_F = 2$ | years | Even unhappy faculty don't leave immediately. From Pavlov and Katsamakas (2020) |
| Time to fill a position | $\tau_h = 2$ | years | From Pavlov and Katsamakas (2020) |
| Faculty attrition | $o_F = (\phi/Q_F)F$ | faculty/year | An outflow from the stock of faculty. We assume a typical turnover rate of faculty $\phi = 0.1$. Poor faculty academic experience increases the turnover rate. |
| Faculty | $F$ | faculty | Initial value is $F_0 = Y_0/\bar{q}_{YF}$ |

**Facilities**

| Variable | Notation | Units | Notes |
|---|---|---|---|
| Student facilities | $B_Y$ | ft² | Available student facilities |
| Panned student facilities | $K_Y$ | ft² | A college may approve construction of new student facilities |
| Faculty office space needed | $\bar{B}_F$ | ft² | Depends on the faculty and required office space per faculty member |
| Faculty facilities that have been constructed | $B_F$ | ft² | The college may have less faculty office space than needed |
| Total facilities | $B = B_Y + B_F$ | ft² | Total student and faculty facilities. This will determine operating and maintenance cost of facilities. |
| Faculty space loading | $\gamma_B$ | dimensionless | A measure of office space availability |
| Percent of approved projects | $\lambda_B = 50$ | percent | We assume that not all construction projects are approved |
| Approved facilities construction | $b_B$ | ft² | New facilities that will be built |
| Student facilities per student | $b_Y = B_Y/Y$ | ft²/student | Available student facilities in per capita terms |
| Basic space per student | $\bar{b}_Y = 100$ | ft²/student | We assume this value includes classrooms, dining, and dorms |
| Office space per faculty member | $b_F = 315$ | ft²/faculty | Office space is typically within the range of 150-350 ft²/person (Brooks, 2022) |
| Construction time | $\tau_B = 3$ | years | Assumed value |
| Competitive edge in facilities | $k$ | dimensionless | |

**Financials**

| Variable | Notation | Units | Notes |
|---|---|---|---|
| Discount rate | $\delta_R$ | dimensionless | Fraction of tuition, room, board, and fees distributed as financial aid |
| Sticker price | $P$ | $/student/year | Tuition, room, board, and fees per student. In the article, we refer to this as "tuition." Assumed initial value is $P_0 = \$31{,}187$. |
| Maximum tuition hike | $\beta = 0.05$ | dimensionless | We assume tuition can increase by no more than 5 percent annually |
| Tuition adjustment delay | $\tau_P = 1$ | years | Assumed value that implies that tuition can be adjusted every year |
| Revenue | $R$ | $/year | Revenue from tuition, room, board and fees and unrestricted gifts |
| Tuition revenue | $R_Y$ | $/year | Revenue must cover expenditures, otherwise college borrows for operations |
| Average net tuition | $NP$ | $/student/year | Tuition less the average financial aid per student |
| Expenditures | $C$ | $/year | All operating costs combined |
| Faculty compensation | $C_F$ | $/year | Proportional to the number of the faculty, $F$, and wage, $\omega$. |
| Debt payments | $C_D$ | $/year | This increases as debt increases |
| Operating and maintenance costs of facilities | $C_B = c_B B$ | $/year | Here $c_B$ is the operating cost of facilities and $B$ is the total facilities. We assume $c_B = 159$ $/ft² |
| Financial aid expenditures | $C_A$ | $/year | The financial aid to students. |
| Net revenue | $S$ | $/year | When positive (negative), referred to as surplus (deficit) |
| Funds borrowed for new construction | $d_B$ | $ | A college either borrows for construction or raises funds through a capital campaign. |
| Cash available for operations | $m_C$ | $/year | Cash available during the fiscal year $\tau_F$ (duration $\tau_F = 1$ year) |
| Accepted endowment draw rate | $a_E = 0.05$ | dimensionless | We assume that a college can withdraw no more than five percent of its endowment per year. Five percent is the 5-year average endowment income and the accepted capital spending practices, according to NACUBO (2020). |

| Available endowment draw | $m_E$ | $/year | Money that can be taken from the endowment fund for operations |
| Funds borrowed for operations | $m_D$ | $/year | This adds to the institutional debt |
| Average faculty compensation | $\omega = 50{,}000$ | $/faculty/year | From Pavlov and Katsamakas (2020) |
| Expenditures per student | $e$ | $/student/year | Used to rank the two colleges |
| Endowment draw | $o_E$ | $/year | Money withdrawn from the endowment fund for operations |
| Unrestricted gifts | $g_u = 0$ | $/year | These gifts could be used for operations. We assume no gifts. |
| Restricted gifts | $g_r = 0$ | $/year | Added to the endowment. We assume no gifts. |
| Cash | $M$ | $ | Initial value is $M_0 = \$0$. From Pavlov and Katsamakas (2020) |
| Debt | $D$ | $ | Initial value is $D_0 = \$0$. From Pavlov and Katsamakas (2020) |
| Endowment | $E$ | $ | Initial value is $E_0 = \$50e6$. From Pavlov and Katsamakas (2020) |

**Ranking agency and applicants**

| *Variable* | *Notation* | *Units* | *Notes* |
|---|---|---|---|
| Applicants pool | $A = 20{,}000$ | students/year | For this model, we assume that the application pool does not change. |
| Ranking index for college $i$ | $I^i$ | dimensionless | Ranking depends on the college's relative reputation, expenditures per student, and student facilities. |

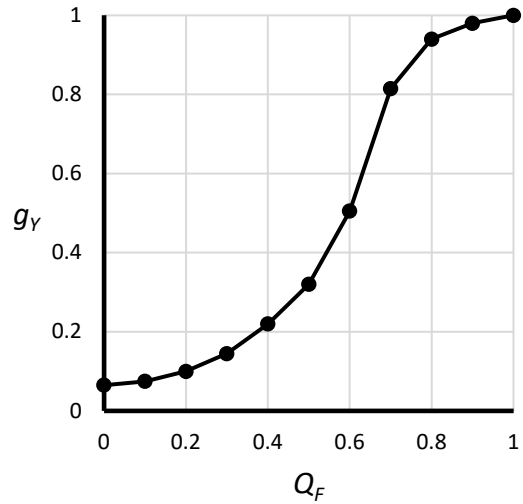

**Figure 13**: Graphical function $g_Y(Q_F)$.

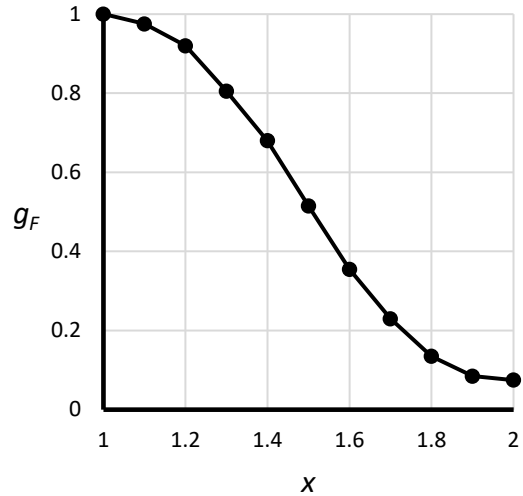

**Figure 14**: Graphical function $g_F(x)$. Here $x = 0.5l_F + 0.5\gamma_B$ is a composite variable consisting of faculty academic load index $l_F$ and faculty space loading $\gamma_B$.

Barnabe, F., 2004, From Ivory Towers to Learning Organizations; the Role of System Dynamics in the Managerialization of Academic Institutions, 22nd System Dynamics International Conference, Oxford.

Baumol, W.J., Bowen, W.G., 1965. On the Performing Arts: The Anatomy of Their Economic Problems. The American Economic Review, 55(1/2), 495-502.

Bell, G., Cooper, M., Kennedy, M., Warwick, J., 2000. The development of the holon planning and costing framework for higher education management. In Proceedings of the 18th system dynamics conference, Bergen, Norway.

Bergland, B.W., Potash, P.J., Heinbokel, J.F., 1999, System Dynamics and Institutional Decision Making, 85th Annual Meeting of the Association of American Colleges and Universities, San Francisco, CA.

Birnbaum, R., 1988. How Colleges Work: The Cybernetics of Academic Organization and Leadership. San Francisco, CA:  John Wiley & Sons.

Bosch, O., Cavana, R.Y. (Eds.), 2018. Systems Education for a Sustainable Planet. MDPI, Basel, Switzerland.

Bowen, H.R., 1980. The Costs of Higher Education: How Much Do Colleges and Universities Spend per Student and How Much Should They Spend? San Francisco, CA:  Jossey-Bass Inc.

Bowen, W.G., 1968. The Economics of the Major Private Research Universities Berkeley, CA:  Carnegie Commission on Higher Education.

Brooks, C., 2022. How to Determine How Much Office Space You Need. Business.com(September 1).

Caskey, J., 2018. The Awkward Economics of Private Liberal Arts Colleges. Cornell University Higher Education Research Institute Working Paper #181.

Cavana, R.Y., 2021. A System Dynamics Scenario Model of the New Zealand Economy: Review and Reflections After 25 Years. In: Cavana, R.Y., Dangerfield, B., Pavlov, O.V., Radzicki, M.J., Wheat, I.D. (Eds.), Feedback Economics: Economic Modeling with System Dynamics. New York:  Springer, 545-582.

Cavana, R.Y., Dangerfield, B., Pavlov, O.V., Radzicki, M.J., Wheat, I.D. (Eds.), 2021. Feedback Economics: Economic Modeling with System Dynamics. Springer, New York.

Davidsen, P., Kopainsky, B., Moxnes, E., Pedercini, M., Wheat, D., 2014. Systems Education at Bergen. Systems, 2(2), 159-167.

Day, R.H., 1974. System simulation: on system dynamics. Behavioral Science, 19(4), 260-276.

Day, R.H., 1981. Dynamic Systems and Epochal Change. In: Sabloff, J.A. (Ed.), Simulations in Archeology. Albuquerque, NM:  University of New Mexico Press, 189-227.

Day, R.H., 1982a. Complex Behavior in System Dynamics Models. Dynamica, 8(2).

Day, R.H., 1982b. Irregular Growth Cycles. American Economic Review, 72, 406-414.

Day, R.H., 1983a. Dynamical Systems Theory and Complicated Economic Behavior. MRG Working Paper #8215. University of Southern California, 161-177.

Day, R.H., 1983b. The emergence of chaos from classical economic growth. The Quarterly Journal of Economics, 98(2), 201-213.

Day, R.H., 1983c. Talk at the International System Dynamics Conference, July 27-30, 1983. MIT System Dynamics Group Memo D-3465.

Day, R.H., 1994. Complex Economic Dynamics, vol. 1.  vol. 1. Boston, MA:  The MIT Press.

Day, R.H., 2001. Complex Economic Dynamics, vol. 2.  vol. 2. Boston, MA:  The MIT Press.

Day, R.H., Koenig, E.F., 1975. On Some Models of World Cataclysm. Land Economics, 51(1), 1-20.

Denning, J.T., 2017. College on the Cheap: Consequences of Community College Tuition Reductions. American Economic Journal: Economic Policy, 9(2), 155-188.

Ehrenberg, R.G., 2000. Tuition Rising: Why College Costs So Much. Cambridge, MA:  Harvard University Press.
34